\documentclass[aip,jcp,superscriptaddress,reprint]{revtex4-1}
\usepackage{epsfig}
\usepackage{color}
\usepackage{amsmath}
\usepackage{amsfonts}
\usepackage{notes2bib}
\usepackage{graphicx}

\newcommand{\erf}{\ensuremath{\mathop{\rm erf}}}

\newcommand{\rb}{\mathbf{r}}
\renewcommand{\sb}{\mathbf{s}}

\newcommand{\kb}{\mathbf{k}}
\newcommand{\avg}[1]{\left<#1\right>}
\newcommand{\len}[1]{\left|#1\right|}
\newcommand{\brac}[1]{\left[#1\right]}
\newcommand{\curly}[1]{\left\{#1\right\}}
\newcommand{\para}[1]{\left(#1\right)}

\DeclareMathOperator{\Tr}{Tr}

\newcommand{\ie}{\emph{i.e.}}
\newcommand{\eg}{\emph{e.g.}}

\newcommand{\rhoq}{\ensuremath{\rho^q}}

\newcommand{\frhoq}{\ensuremath{\hat{\rho}^q}}

\newcommand{\chiqq}{\ensuremath{\chi^{qq}}}
\newcommand{\fchiqq}{\ensuremath{\hat{\chi}^{qq}}}

\newcommand{\fchiqqo}{\ensuremath{\hat{\chi}^{(0)qq}}}
\newcommand{\fchiqqt}{\ensuremath{\hat{\chi}^{(2)qq}}}

\newcommand{\rhob}{\rho_{\rm B}}

\newcommand{\sLJ}{\ensuremath{\sigma_{\rm LJ}}}

\newcommand{\Qb}{\ensuremath{\mathcal{Q}}}

\newcommand{\Rbar}{\overline{\textbf{R}}}

\newcommand{\sgn}{\ensuremath{{\rm sgn}}}

\newcommand{\phiB}{\ensuremath{v^q_{\rm B}}}
\newcommand{\phiD}{\ensuremath{\Delta v^q_{\rm D}}}
\newcommand{\phiC}{\ensuremath{\Delta v^q_{\rm C}}}

\newcommand{\V}{\ensuremath{{v^q}}}

\begin{document}

\title{The Influence of Distant Boundaries on the Solvation of Charged Particles}

\author{Richard C. Remsing}
\email[]{rremsing@temple.edu}
\affiliation{Institute for Computational Molecular Science and Department of Chemistry,
Temple University, Philadelphia, PA 19122}

\author{John D. Weeks}
\email[]{jdw@umd.edu}
\affiliation{Institute for Physical Science and Technology
		 and Department of Chemistry and Biochemistry,
		 University of Maryland, College Park, MD 20742}

\begin{abstract}
The long-ranged nature of the Coulomb potential requires a proper accounting for the influence of even distant electrostatic boundaries in the determination of the solvation free energy of a charged solute. 
We introduce an exact rewriting of the free energy change upon charging a solute that explicitly isolates the contribution
from these boundaries and quantifies the impact of the different boundaries on the free energy.
We demonstrate the importance and advantages of appropriately referencing the electrostatic potential to that of the vacuum through
the study of several simple model charge distributions, for which we can isolate an analytic contribution from the boundaries that can be
readily evaluated in computer simulations of molecular systems. 
Finally, we highlight that the constant potential of the bulk dielectric phase --- the Bethe potential --- cannot
contribute to the solvation thermodynamics
of a single charged solute when the charge distributions of the solvent and solute do not overlap in relevant configurations.
But when the charge distribution of a single solute can overlap with the intramolecular charge
distribution of solvent molecules, as is the case in electron holography, for example, the Bethe potential is needed when comparing to experiment.  Our work may also provide insight into the validity of ``extra thermodynamic assumptions'' traditionally made during
the experimental determination of single ion solvation free energies.
\end{abstract}

\maketitle

\section{Introduction}
The solvation of ions and other charged particles is of fundamental importance in chemical and biological processes,
often influencing the solvation, self-assembly, and chemistry of macromolecules and materials~\cite{Collins:2012,HofmeisterReview,England:2011rc,Canchi:2013nx,Dang_ChemRev_2006,Knipping_Science,Geissler:2013rz,Tobias:2013ty,Netz:2012fk,Remsing:2015ab,Remsing:JACS:2017,Remsing:CPL:2017}.
However, the long-ranged nature of electrostatic interactions
requires properly accounting for the influence of even distant boundaries,
and their possible contributions to solvation thermodynamic properties
has long been a source of contention
~\cite{Bethe_1928,HarrisBook,Kholopov:2006,Wilson:1989,Pratt:1992,Hummer:1996,Leung:2007,Harder:2008,Lueng:2009,Baer:2012,Bardhan:2012,Shi:2013,Beck:2013,Reif_2016,Duignan2017,Pratt_2017,Beck_2018,Duignan_2018,Hunenberger_2018,Pratt_2018}.

Contributions from distant boundaries to the free energy of charging a solute are linear in the charge of the solute, $Q$,
such that these terms cancel for neutral ion combinations.
Thus, distant boundaries have no influence on most experimental solubility measurements,
and more generally on solutions that satisfy electroneutrality.
However, system boundaries do need to be properly taken into account when decomposing experimental free energies
into ion-specific components, as well as in the computation of single ion solvation free energies~\cite{Baer:2012,Shi:2013,Beck:2013,Duignan2017,Remsing_JPCL_2014,Kathmann:2011,Kastenholz:2006qf,Kastenholz:2006nx,Hunenberger2011,Duignan2017a}, where differing empirical ``extra-thermodynamic'' assumptions are often introduced.

Early work identified the importance of electrostatic potential differences across phase boundaries in
solvation thermodynamics~\cite{Pratt:1992},
and pioneering studies on ion solvation showed that the asymmetry of aqueous ion solvation --- anions are more favorably
hydrated than cations --- arises from a non-zero electrostatic potential inside uncharged solute cores~\cite{Hummer:1996,Ashbaugh:2000,Rajamani:2004}.
As we discuss throughout this work, the potential difference across a solute core, and any physical boundary in general,
consists of a structural component, 
as well as a constant electrostatic potential of the bulk phase,
the so-called Bethe potential~\cite{Bethe_1928,HarrisBook,Kholopov:2006}.
Many of these studies recognized the importance of referencing appropriately to the bulk~\cite{Hummer:1996,Ashbaugh:2000,Rajamani:2004}, removing the Bethe potential contribution through the use
of Ewald summation to compute the potential~\cite{HarrisBook}.
However, most of these studies did not include a description of distant boundaries, such that electrostatic potentials
inside the cavity were not referenced to vacuum, and yielded potentials that were opposite in sign to the expected value;
positive instead of negative for SPC/E water, for example~\cite{Ashbaugh:2000,Rajamani:2004}.
Recent work has clarified this issue and we believe the community is coming to a consensus regarding the impact of distant boundaries, such
that their effects need to be included to yield accurate single ion thermodynamics~\cite{Leung:2007,Harder:2008,Lueng:2009,Baer:2012,Shi:2013,Beck:2013,Reif_2016,Duignan2017,Remsing_JPCL_2014,Kathmann:2011,Kastenholz:2006qf,Kastenholz:2006nx,Hunenberger2011,Duignan2017a,CoxGeissler}.

In this work, we use a combination of molecular simulation and analytic models to isolate and directly focus on contributions from distant boundaries or different boundary conditions affecting the free energy of charging an ionic solute in typical classical solvent models with point charges
embedded in repulsive molecular cores, studying in particular the influence of the Bethe potential~\cite{Bethe_1928,HarrisBook,Kholopov:2006}.
We start from an exact decomposition of the charging free energy into preexisting boundary terms, where solvent perturbations exist independent of the ion charge, and terms arising from local structural perturbations induced by the ion charge. We discuss three particularly relevant classes of distant boundaries to be considered in molecular simulations: i) site-based periodic boundaries, ii) a new class of hypothetical boundaries especially useful for determining the Bethe potential, which we refer to as Bethe boundaries, and iii) realistic structural boundaries, such as repulsive walls or liquid-vapor interfaces. 
By focusing on the \emph{structural} origins of potential differences arising from distant boundaries, instead
of just the potentials themselves, we hope to clarify subtleties that can arise in the computation of ionic charging free energies.

As in earlier work~\cite{Remsing_JPCB_2016}, we first study the charging free energy of a Gaussian charge distribution, where the total ion charge $Q$ is smeared out as a Gaussian on a molecular length scale.
This model ionic charge distribution can overlap with the point charges in nearby solvent molecules, but the solvent responds linearly to the smeared Gaussian charge to a very good approximation.
This linear response allows us to compute simple but very accurate analytic expressions for the charging free energy of Gaussian charges in dielectric solvents for each class of distant boundary conditions, illustrating their different effects. The additional contributions arising from local electrostatic boundaries induced by repulsive molecular cores in realistic ion models will be discussed later.

Physically, Gaussian charges may serve as idealized models for the solvation thermodynamics of quantum charge distributions, such as electrons, that can probe intramolecular charge distributions in molecular systems~\cite{Kathmann:2011,Berne:AnnRevPhysChem:1986,Sprik:CompPhysRep:1988,Chandler:AnnuRevPhysChem:1994,Bischak_2017,Remsing:PNAS:2018}.
For this class of solute models, in which the charge density of the solute and solvent can overlap, the Bethe potential can indeed
contribute to solvation thermodynamics. 

In contrast, most realistic molecular ion models have a point charge embedded within an excluded volume core that prevents significant overlap of the ion charge distribution
with that of the solvent molecules.
For such models, we show that the solvation free energies of single ions are independent of the constant Bethe potential.
We further emphasize this point through the study of atomistic models with smeared intramolecular charge distributions,
where there is a complete decoupling of changes in the intermolecular interactions and the Bethe potential.
We then conclude with a discussion of the impact of our models for understanding boundary effects
in the estimation of ion solvation thermodynamics for realistic systems.

\section{Separating Distant Boundaries from Local Solvation}

The thermodynamic implications of distant boundaries can be extracted from an exact formulation \cite{Remsing_JPCB_2016} of the total free
energy change generated by a modified solute-solvent Coulomb interaction energy $\Psi_\lambda(\Rbar)$ in the system Hamiltonian between a partially charged model ion with charge $\lambda Q$ and the solvent with
$N_{\rm C}$ full charges at sites $\rb_i(\Rbar)$ in a configuration $\Rbar$ given by
\begin{equation}
\Psi_\lambda(\Rbar)= \int d\rb \int d\rb' \frac{\rho^Q_\lambda(\rb') \rhoq(\rb;\Rbar)}{\len{\rb-\rb'}}
\end{equation} 
as the linear coupling parameter $\lambda$ is varied between zero and one.
Here
\begin{equation}
\rho^{q}(\rb;\Rbar)= \sum_{i=1}^{N_{\rm C}} q_i \delta (\rb-\rb_i(\Rbar))
\end{equation}
is the configurational charge density of the solvent, denoted by the superscript $q$.
We focus on model ion charge distributions given by
 \begin{equation} 
 \rho^Q_\lambda(\rb) = \lambda Q \rho_{G}(r; l), 
 \label{eq:rhog}
 \end{equation}
where $\rho_{G}(r; l)$ is a normalized Gaussian distribution with width $l$,
\begin{equation}
\rho_{\rm G}(r;l)\equiv \frac{1}{l^3\pi^{3/2}}e^{-r^2/l^2}.
\end{equation}
In the smeared Gaussian charge model discussed below, $l$ is a molecular length scale;  the usual $\delta$-function point charge model arises in the limit as $l \to$ 0.
Note that $\Psi_\lambda(\Rbar)$ is the only term in the solute-solvent Hamiltonian that depends explicitly on $\lambda$.

Following standard coupling parameter methods we differentiate the partition function determining the free energy $G_{\lambda}^{c}(Q)$ of the partially coupled system with respect to $\lambda$ and integrate over $\lambda$. This gives an exact expression for the charging free energy
$\Delta G^{c}(Q) = G_{1}^{c}(Q) -G_{0}^{c}(Q)$ that has an especially simple form when written in terms of the well-defined solute and solvent charge distributions:
\begin{equation}
\Delta G^{c}(Q)= \int_0^1 d\lambda \int d\rb \int d\rb' \frac{\rho^Q (\rb') \rhoq _{\lambda}(\rb)}{\len{\rb-\rb'}}.
\label{eq:cfe}
\end{equation}
Here
\begin{equation}
 \rho^q_\lambda(\rb)=\avg{\rho^q(\rb;\Rbar)}_\lambda
\end{equation}
is the charge density of the solvent,
where
$\avg{\cdots}_\lambda$ denotes a normalized ensemble average in state $\lambda$,
and $\rho^Q (\rb')=\rho^Q _{\lambda=1} (\rb')$ is the charge density of the fully charged solute.

Consistent with Green's reciprocity relation~\cite{Zangwill},
$\Delta G^{c}(Q)$ can be written in two mathematically equivalent forms involving the solute or solvent electrostatic potential, as used in earlier work, but each expression suggests different procedures for their evaluation in theory and molecular simulations. Most earlier work~\cite{Hummer:1996,Rajamani:2004,Bardhan:2012} considered the  $\lambda$-averaged electrostatic potential of the solvent
\begin{equation}
\overline{v}^{q} (\rb')= \int_0^1 d\lambda v^q_\lambda(\rb') =\int_0^1 d\lambda \int d\rb  \frac{\rhoq _{\lambda}(\rb)}{\len{\rb-\rb'}},
\end{equation}
and the case where the solute charge density is a
point charge $\rho^Q (\rb')=Q\delta (\rb')$, embedded within an excluded volume core at the origin.
Eq.~\ref{eq:cfe} then gives
 \begin{equation}
 \label{eq:vbarq0}
\Delta G^{c}(Q)= Q\overline{v}^{q} (0).
\end{equation}

However, despite its simple form requiring only the value of the solvent electrostatic potential at the center of the ion core, this expression can cause confusion regarding the influence of boundaries, because, unlike the solvent charge density, the solvent electrostatic potential is not defined until boundary conditions are specified~\cite{HarrisBook,Kholopov:2006,Kleinman_PRB_1981,Euwema_JPCS_1975,Makov_PRB_1995}. These boundary terms appear only implicitly in Eq.~\ref{eq:vbarq0} and
typically lead to constant shifts in the electrostatic potential that can significantly alter the value of
$\overline{v}^{q} (0)$ and the resulting charging free energy.

As in our previous work~\cite{Remsing_JPCB_2016} we consider Gaussian as well as point charge distributions and analyze instead the alternative form of Eq.~\ref{eq:cfe}
that relates the well-defined solute electrostatic potential
\begin{equation}
v^Q(\rb)=  \int d\rb' \frac{\rho^Q (\rb')}{\len{\rb-\rb'}},
\end{equation}
to the solvent charge density. In this rewriting of Eq.~\ref{eq:cfe} solvent boundary perturbation terms appear explicitly and can be readily identified and analyzed:
\begin{align}
\Delta G^c(Q)
&=  \int d\rb v^Q(\rb) \int_0^1 d\lambda \Delta \rhoq_\lambda(\rb) +\int d\rb v^Q(\rb) \rhoq_0(\rb) \label{eq:charging2} \\
&\equiv  \Delta G^c_{\rm IS}(Q)+\Delta G^c_{\rm PB}(Q).
\label{eq:charging}
\end{align}
Here $\Delta \rhoq_\lambda(\rb)= \rhoq_\lambda(\rb) - \rhoq_0(\rb)$.

The first term in Eq.~\ref{eq:charging}, $\Delta G^c_{\rm IS}(Q)$, corresponds to the portion of the charging free energy that arises
from \emph{induced structural changes} in the solvent generated by the charging process, as emphasized in previous work~\cite{Remsing_JPCL_2014,Remsing_JPCB_2016,Remsing_JPCB_2018}.
In this paper, we focus on the second component of the charging free energy, $\Delta G^c_{\rm PB}(Q)$ given by the last term in Eq.~\ref{eq:charging2}.
This part of the free energy arises from \emph{preexisting boundaries} in the system that lead to electrostatic inhomogeneities
even in the absence of any solute charge; hence, it involves $\rho^q_0(\rb)$.

When a boundary is very far from the solute, the nonuniform structure near the boundary is unaffected by the solute  charging process,
and the thermodynamic impact of the boundary is restricted to $\Delta G^c_{\rm PB}(Q)$.
Much of our discussion will concentrate on such preexisting \emph{distant} boundaries.
However, nonuniformities are also induced by realistic solute cores even in the absence of a charge,
and these \emph{local} boundaries will influence both $\Delta G^c_{\rm IS}(Q)$ and $\Delta G^c_{\rm PB}(Q)$.
Solute core boundaries are present in physical models of ions, and we will discuss the influence
of such boundaries when appropriate.
%

\section{Types of System Boundaries}
We group the distant system boundaries into three classes, accessible in computer simulations, in order to best
illustrate their relative roles in solvation thermodynamics.

\subsection{Site-Based Periodic Boundaries}
The first class of boundaries that we consider are \emph{periodic boundary conditions}
defined according to atomic sites (PBCs),
which are typically used in molecular simulation~\cite{CompSimLiqs}. 
We consider a periodic array of cubic simulation cells of length $L$,  to mimic an infinite system.
The periodicity is defined on an intramolecular site basis,
such that a single site crossing the boundary moves to the other side of the simulation cell.
In the notation of H\"{u}nenberger and others, this is referred to as ``P-type'' periodic boundaries~\cite{Hummer:1996,Duignan2017,Kastenholz:2006qf,Kastenholz:2006nx,Hunenberger2011,Duignan2017a,Hummer:1997c,Hummer:1997,Hummer:1998b,Hummer:1998c}.
This is in contrast to ``M-type'' periodic boundaries, which are defined on a molecular basis.
These
keep molecules intact within the simulation cell~\cite{Hummer:1996,Duignan2017,Kastenholz:2006qf,Kastenholz:2006nx,Hunenberger2011,Duignan2017a,Hummer:1997c,Hummer:1997,Hummer:1998b,Hummer:1998c}
and are typically only useful for small molecules.
An infinite system, as often considered in dielectric continuum theories, is obtained in the limit $L\rightarrow\infty$. 
Site-based PBCs have an almost trivial impact on thermodynamics since they make no contribution to $\rhoq_0(\rb)$.
Periodic boundaries result in an ionic charging free energy that arises purely from local structural perturbations,
both charging induced, $\Delta G^c_{\rm IS}$, and from any local core contributions to $\rhoq_0(\rb)$
that may contribute to $\Delta G^c_{\rm PB}$.
Note that PBCs may lead to additional finite size effects on $\Delta G^c_{\rm IS}$
that have been discussed throughout the literature~\cite{Remsing_JPCB_2018,Hummer:1997,Figueirido:1995,CoxGeissler},
and we ignore these well understood corrections.%

\begin{figure*} 
\begin{center}
\includegraphics[width=\textwidth]{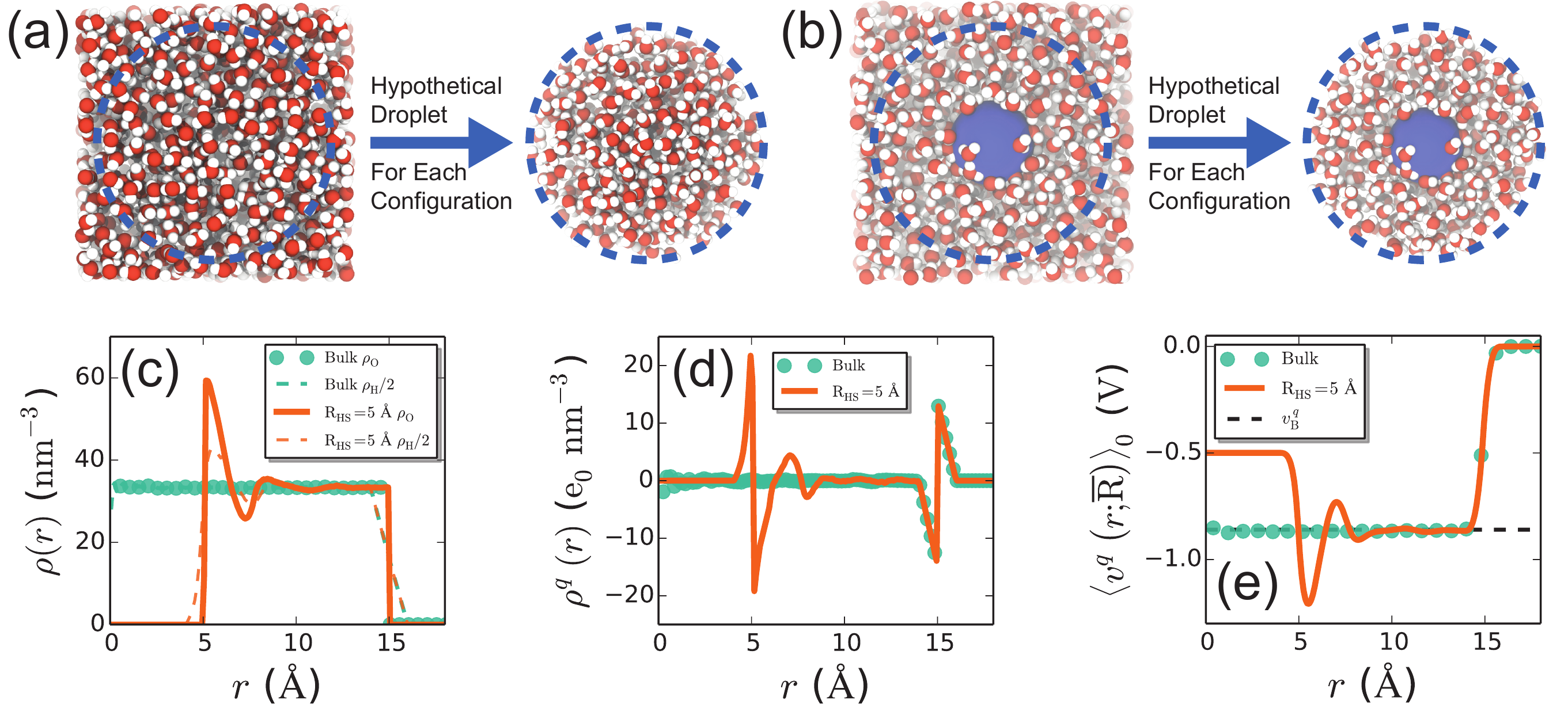}
\end{center}
\caption[Calculating absolute electrostatic potentials]
{(a) Schematic depiction of carving a hypothetical (Bethe) droplet of
neutral molecular charge densities out of a configuration of bulk water.
The artificial droplet boundary
is chosen to have a radius of 15~\AA \ (blue dashed line), beyond which
water molecules are omitted from the calculation of electrostatic properties.
Here, this boundary is defined according to the oxygen site of the water molecule
and bonds are not broken; an intact water molecule is within the droplet
if its oxygen site is within the boundary, and it is removed from consideration otherwise.
(b) Schematic illustration of carving out a Bethe droplet of radius 15~\AA \ in a system
with a hard sphere-like solute (blue surface)
with a water-excluded radius (${\rm R_{HS}}$) of roughly 5~\AA \ centered at the origin.
Lower panels show the resulting (c) oxygen and hydrogen densities, with $\rho_{\rm H}(r)$ scaled by a factor of two,
(d) charge densities,
and (e) electrostatic potentials determined by integrating Poisson's equation. 
The potential $\phiB$ obtained from Equation~\ref{eq:bethemol} is also shown
as a dashed line.
Simulation details can be found in our previous work~\cite{Remsing_JPCL_2014}.
}
\label{fig:droplet}
\end{figure*}

\subsection{Bethe Boundaries}
The second class of boundaries is also hypothetical, and
can be used as a tool in computer simulations to calculate the average potential, $\phiB$, of a bulk dielectric phase relative to the vacuum from structural considerations alone, termed the Bethe potential.

The original development of the Bethe potential was concerned with finding
the mean inner potential at the center of an infinite crystal lattice~\cite{Bethe_1928,HarrisBook}.
Additional work on this issue led to the insight that the potential of a bulk phase is arbitrary
\emph{until boundary conditions are defined}~\cite{HarrisBook,Kholopov:2006,Kathmann:2011,Kleinman_PRB_1981,Euwema_JPCS_1975,Makov_PRB_1995}. 
If an ideal bulk phase is defined as the $L\rightarrow\infty$ limit of site-based PBCs, the limiting potential of the bulk phase is zero.
However, real systems are confined within a finite volume defined by a set of boundaries whose effects must be considered.

We define the Bethe boundary to act on a specific molecular coordinate, such as the center of mass or an atomic position.
Here, we focus on the oxygen atom of a water molecule. 
When the oxygen site of a molecule in a particular configuration is outside this idealized boundary in a molecular simulation,
the molecule's contribution to the number density of the solvent is set to zero.
Solvent molecules within the boundary remain intact and feel the same intermolecular forces as if the boundary were not there.
Thus, the average density of the solvent changes in a step function manner across the boundary from $\rhob$ to zero, in close analogy to the concept of a Gibbs dividing surface in interfacial physics where a uniform bulk is supposed to extend unchanged up to the dividing surface~\cite{WidomBook}.

We choose a spherical droplet boundary for simplicity, analogous to what is often imagined in classical electrostatic discussions
of dielectric media~\cite{Zangwill,Jackson}.
Here the average potential is one-dimensional by symmetry --
but one could alternatively consider Bethe cubic, monoclinic, and other cells~\cite{Bethe_1928,HarrisBook,Kholopov:2006}. 
The process for computing the Bethe potential in this manner is illustrated in Figure~\ref{fig:droplet}a,b.
There, we show snapshots from simulations of water in (a) the bulk
and (b) with a hard sphere-like solute at the origin (blue surface) and illustrate the process
of carving out a hypothetical spherical droplet in each configuration to obtain averages.
The change in the average electrostatic potential across this boundary is the Bethe potential.
Therefore, we refer to this class of boundaries as \emph{Bethe boundaries}.
Despite the spherical geometry of the boundary shown in Figure~\ref{fig:droplet},
there is no curvature dependence to the potential difference across a Bethe boundary,
because the solvent does not respond to its presence. A physical boundary will indeed
display a curvature dependence in the potential difference across it~\cite{Palmeri:2013}.

Because the Bethe boundary is defined to act on oxygen atoms and maintains neutral molecules within the droplet,
the resulting nonuniform oxygen density, $\rho_{\rm O}(r)$, is a step function at the Bethe boundary in both systems, Figure~\ref{fig:droplet}c.
The hydrogen density, $\rho_{\rm H}(r)$, smoothly transitions from the bulk to zero across the boundary
over a distance of $2r_{\rm OH}$, where $r_{\rm OH}$ is the O-H bond length.
The resulting differences in the oxygen and hydrogen densities create a non-zero charge density, $\rhoq(r)$,
in the vicinity of the boundary, Figure~\ref{fig:droplet}d.

Around a spherical nonpolar solute that excludes water from a radius of roughly 5~\AA, 
a small fraction of hydrogen bonds are broken at the surface (roughly 3.6/molecule in bulk to 3.2/molecule). 
Water molecules reorient to point the resulting free OH groups toward the solute surface.
In addition to the layering of water at the surface, this reorientation results in non-trivial
density correlations near the solute due to interfacial structural response, Figure~\ref{fig:droplet}c.
Structural ordering at the surface of a solute results in additional oscillations of the charge density near
the solute surface, as shown in Figure~\ref{fig:droplet}d.
Once this droplet is defined, Poisson's equation is integrated to yield the electrostatic potential
in configuration $\Rbar$ with respect to the vacuum~\cite{Zangwill,Jackson},
\begin{align}
v^q(r;\Rbar)
&= \int_r^\infty \frac{dr'}{r'^2} \int_0^{r'} dr'' r''^2 \rhoq(r;\Rbar)
\end{align}
where the configurational potential is referenced to zero in the vacuum region located at $r\gg r_c$.
The average electrostatic potential is then obtained as
the ensemble average over configurations.
The average potential is equal to $\phiB$ within the bulk and equal to zero well outside the Bethe boundary.
This allows for the determination of $\phiB$ as the difference in the electrostatic potential between these two limits,
Figure~\ref{fig:droplet}e.
This method of determining 
$\avg{v^q(r;\Rbar)}_\lambda$ explicitly references the average electrostatic potential to 
its value in vacuum, but in contrast to earlier work does not require the presence of a physical phase boundary
during the simulation~\cite{Baer:2012,Duignan2017,Remsing_JPCL_2014,Kathmann:2011,Duignan2017a}.

This approach can readily be extended to determine
electrostatic potentials in nonuniform systems.
Indeed, the potential determined in the presence of a cavity,
also shown in Figure~\ref{fig:droplet}e, yields $\phiB$ as the potential in the bulk phase
and also allows for the estimation of $\phiC$, the change in the electrostatic potential
due to the presence of the cavity, also appropriately referenced to vacuum.
This cavity potential depends on both the size of the solute core and the model of water
used in a non-trivial manner~\cite{Remsing_JPCL_2014,Ashbaugh:2000},
but the general Bethe boundary formalism is not limited to any particular solute or solvent interaction
potential.
 
With the implicit assumption of Bethe boundaries, earlier work has shown that
the Bethe potential for rigid molecular models can be written as~\cite{Wilson:1989,Pratt:1992,Harder:2008,Remsing_JPCL_2014}
\begin{equation}
\phiB=-\frac{4\pi}{3}\rhob \Tr\curly{\Qb_{\rm mol}},
\label{eq:bethemol}
\end{equation}
where $\rhob$ is the bulk density and $\Tr\curly{\Qb_{\rm mol}}$ is the trace of the primitive quadrupole tensor of a single
molecule, $\Qb_{\rm mol}$. 
More complex expressions must be used for models whose charge distribution can fluctuate~\cite{Remsing_JPCL_2014,Kathmann:2011}.
The value of the Bethe potential determined from Eq.~\ref{eq:bethemol} for SPC/E water is shown as a dashed line in
Fig.~\ref{fig:droplet}e and agrees with that predicted from the Bethe droplet approach. 

Equation~\ref{eq:bethemol} and its generalizations~\cite{Remsing_JPCL_2014,Kathmann:2011}
illustrate that the Bethe potential is tied to the trace of the quadrupole moment (or spherical second moment)
of polar molecules.
Our Bethe droplet approach offers an equivalent but alternative structural perspective
to build intuition about the electrostatics of dielectric media.
Here the Bethe potential is shown to arise from the molecular
structure and non-zero charge density at a Bethe boundary.

We conclude this section by noting that
the Bethe potential of dipolar molecules is well known to depend on the choice of molecular center~\cite{Duignan2017,Hummer:1997c,Hummer:1998c}.
In our approach, this center-dependence immediately manifests itself
in the structure of the nonuniform charge density induced near a Bethe boundary.
As an example, consider defining the Bethe boundary for water to act on hydrogen sites, instead of the oxygen site as used above.
The oxygen sites can penetrate the boundary, instead of hydrogen.
The resulting charge density will be negative outside of the boundary, inverting the Bethe potential.
In contrast, quadrupolar molecules, such as a rigid model of methane (CH$_4$),
do not show such a site dependence, because of their molecular symmetry~\cite{Harder:2008}.
In this case, a Bethe boundary located at $r$ that acts on H,
is equivalent to placing a Bethe boundary that acts on C at $r-r_{\rm CH}$, where $r_{\rm CH}$ is the C-H bond length.

\subsection{Distant Structural Boundaries}
The final class of boundaries we consider are \emph{distant structural boundaries}, which arise
from physical interfaces present in the system that are far from the solute in the context of bulk solvation.
These boundaries separate the solvent from a true \textit{vacuum}.
In the case of water at ambient conditions, the vacuum is often approximated by the low-density vapor phase,
such that the distant boundary is an aqueous liquid-vapor interface.
In situations where the vacuum is not well approximated within a molecular simulation
by the vapor phase, which may have its own Bethe potential,
hard walls can be used to create a distant structural boundary between vacuum and the liquid or vapor phase.

The solvent can respond to the presence of distant structural boundaries, creating structural inhomogeneities.
Such structural perturbations lead to non-trivial density and charge density correlations near the boundary.
These correlations generate a non-zero electrostatic potential difference across the boundary.

Creating a physical interface can be imagined as a two-step process.
In the first step, we create an ideal Bethe boundary at the desired location of the physical interface.
This boundary creates a potential difference of $\phiB$.
In the second step, the physical interface is created by introducing any walls and/or external potentials
and allowing the Bethe boundary configurations to relax at the interface, for example, by reorientation of solvent molecules.
The solvent response involved in this second step modifies the electrostatic potential difference
across the boundary by an amount $\phiD$.
Thus, the total electrostatic potential difference across a physical, structural boundary
is
\begin{equation}
\Delta v^q = \phiB+\phiD.
\end{equation}
For rigid, and non-polarizable molecules,
such as the extended simple point charge (SPC/E) model of water,
$\phiD$ across a planar interface
is exclusively due to molecular dipoles~\cite{Wilson:1989,Pratt:1992,Palmeri:2013,Remsing_JPCB_2015},
hence the subscript ${\rm D}$,
which will depend on the choice of molecular center used
to bin the dipole or polarization density underlying its computation~\cite{Duignan2017,Hummer:1997c,Hummer:1998c},
but this dependence vanishes when considering an additional boundary, like that of an excluded volume solute core,
as discussed further below.
In general, this potential difference can
arise from higher order molecular multipoles as well,
depending on the geometry of the boundary and the polarizability of the model~\cite{Palmeri:2013}.

The separation of Bethe and structural components of the potential within the
two-step process for creating a distant structural boundary can be leveraged to 
enhance the efficiency of simulation estimates of ion solvation free energies.
The Bethe droplet approach allows for the determination of the cavity potential without the need for explicit physical boundaries as a reference during the simulation,
\eg \ by inclusion of a liquid-vapor interface, as is usually done in most current approaches~\cite{Baer:2012,Duignan2017,Remsing_JPCL_2014,Kathmann:2011,Duignan2017a}.
The inclusion of such an interface, along with a solute in bulk solution, requires
large system sizes that become computationally prohibitive in \emph{ab initio} simulations.

By using a Bethe boundary to reference the potential to vacuum, one can simulate a small system consisting
of a solute in bulk and compute $\phiC$.
The effects of distant structural boundaries can be studied separately, focusing on that interface alone,
and incorporated into the potential estimate in a second step.
Note that the same convention for defining the molecular center must be used in both
steps of this process in order to yield consistent values for the potential components.
This method of treating distant boundaries can significantly reduce the system sizes needed
and, therefore, the computational cost required to compute \emph{ab initio} estimates of ion solvation thermodynamics by avoiding the need for including an explicit interface in the simulated system~\cite{Baer:2012,Duignan2017,Kathmann:2011,Duignan2017a}.
Moreover, this approach readily enables the study of ion hydration \emph{away from liquid-vapor coexistence
through changing the temperature and pressure (or volume) of the bulk solvent.}

\section{Boundary Effects on the Solvation of a Gaussian Charge Distribution}

We now illustrate the major effects of boundary conditions through solvation of Gaussian charge distributions,
for which accurate theoretical and computational frameworks were previously developed~\cite{Remsing_JPCB_2016}.
In that work, it was shown that analytical linear response theory expressions accurately
describe the insertion of a Gaussian charge distribution into a dielectric fluid
for reasonably small magnitudes of the total charge when the width of the Gaussian is chosen on the scale of typical nearest-neighbor distances in the fluid.
For classical models of water, widths of $l>3$~\AA \ and magnitudes $Q<4e_0$ suffice for linear response theory to hold.
We proceed by examining how the three types of boundary conditions modify the free energy of increasing the magnitude of
a Gaussian charge distribution from 0 to $Q$ in a neutral dielectric solvent.
Because the response of the solvent to this charge distribution is linear,
there are no subtleties due to asymmetries with respect to $Q$
or other non-linear responses that typically plague analytic treatments
for inserting a point charge within an excluded volume~\cite{Hummer:1996,Bardhan:2012,Remsing_JPCB_2016,Rajamani:2004}.
Moreover, the absence of an excluded volume core also enables us to focus entirely on effects from non-local, distant boundaries,
whereas the presence of a solute core would require
additional complexities to separate the effects of local structural rearrangements arising from this strong solute-solvent interaction.

The simplifications afforded by the use of Gaussian charges permit an analytic treatment of distant boundaries and these same effects will be observed with ``real'' ions, with additional local contributions arising from the presence of (core) boundaries and non-linear and asymmetric local solvent responses to charging~\cite{Duignan2017,Duignan2017a,Remsing_JPCB_2016}.
We first generalize the treatment of Gaussian charge solvation
developed previously~\cite{Remsing_JPCB_2016}, before explicitly treating the various boundaries.
In the case of linear Gaussian charging, the solute charge density at a coupling parameter $\lambda$
is given by Eq.~\ref{eq:rhog}.
The potential arising from this charge distribution is readily determined using
\begin{equation}
v^{Q}_{\lambda}(\rb) = \int d\rb' \frac{\rho^{Q}_{\lambda}(\rb)}{\len{\rb-\rb'}}=\lambda Q \frac{\erf(r/l)}{r}.
\end{equation}
To determine a linear response approximation to the exact charging free energy, and its components
as given by Eq.~\ref{eq:charging},
we need to develop an expression for the change in solvent charge density induced by the charging process.
We follow our previous work~\cite{Remsing_JPCB_2016}, but consider a non-zero $\rhoq_0(\rb)$ for generality.
In this case, $\rhoq_0(\rb)$ accounts for the nature of preexisting boundaries in the system, both distant and local.
To linear order in the solute potential, the induced solvent response is given by
\begin{equation}
\Delta\frhoq_\lambda(\kb)\approx -\beta\fchiqq_0(\kb)\hat{v}^Q_\lambda(\kb),
\end{equation}
where $\hat{f}$ indicates the Fourier transform of $f$ and
$\chiqq_0(\len{\rb-\rb'})$ is the charge-charge linear response function of the bulk solvent.

The Gaussian nature of $\hat{v}^Q_\lambda(\kb) =(4\pi \lambda Q/k^{2}) e^{-(kl)^{2}/4}$
cuts off the large $k$ components in the density response,
so that an expansion of the linear response function to second order in $k$ is valid for sufficiently large $l$,
\begin{equation}
\fchiqq_0(\kb)\sim \fchiqqo_0+k^2\fchiqqt_0.
\end{equation}
For a neutral solvent, $\fchiqqo_0=0$ due to electroneutrality~\cite{StillLove1,StillLove2,StillLove}.
The second moment of the charge-charge linear response function is related to the dielectric constant
through a generalization of the Stillinger-Lovett moment conditions~\cite{StillLove1,StillLove2,StillLove,TruncCoul},
$4\pi\beta\fchiqqt_0 = 1-1/\epsilon$.
Then the induced solvent charge density response is
\begin{equation}
\Delta\rhoq_\lambda(r)=-\lambda Q\para{1-\frac{1}{\epsilon}}\rho_G(r).
\label{eq:response}
\end{equation}
Although this charging-induced response is independent of the boundaries,
the total charge density is not, $\rhoq_\lambda(\rb)=\Delta \rhoq_\lambda(r)+\rhoq_0(\rb)$.
It is this difference in the total charge density that gives rise to boundary effects on solvation thermodynamics.
The charging free energy can then be obtained by insertion of Eq.~\ref{eq:response} into Eq.~\ref{eq:charging},
which yields
\begin{align}
\Delta G^c(Q) &= \Delta G^c_{\rm IS}(Q) + \Delta G^c_{\rm PB}(Q) \\
&= -\frac{Q^2}{l\sqrt{2\pi}}\para{1-\frac{1}{\epsilon}} + \Delta G^c_{\rm PB}(Q) ,
\label{eq:genFE}
\end{align}
where $\Delta G^c_{\rm PB}(Q)$ depends on the nature of the specific preexisting boundaries through $\rhoq_0(\rb)$.
Equation~\ref{eq:genFE} shows that the contribution to the charging free energy from changes in solvent structure
induced by the charging process is general and independent of the system boundaries.
We now examine $\Delta G^c_{\rm PB}(Q)$ for the three types of boundaries described in the previous section.

\subsection{Periodic Boundary Conditions}
We previously treated the solvation of Gaussian charge distributions under PBCs
and their infinite size limit ($L\rightarrow\infty$)~\cite{Remsing_JPCB_2016}.
Under these conditions, the average solvent charge density in the absence of
a solute is zero everywhere, $\rhoq_0(\rb)=0$.
Therefore, $\Delta G^c(Q)=\Delta G^c_{\rm IS}(Q)$
and $\Delta G^c_{\rm PB}(Q)=0$.
Thus, if one assumes PBCs with Gaussian charges, the charging free energy
is due entirely to the solvent response and there is no contribution from a pre-existing charge density
arising from nonelectrostatic boundaries in the system.
The resulting free energies are parabolic and symmetry with respect to $Q$, as shown in Fig.~\ref{fig:fes}a.
We emphasize that this is true when $\rhoq_0(\rb)=0$.
A solute with a physical excluded volume
core will have $\rhoq_0(\rb)\ne 0$ due to the core boundary.
Note, however, that PBCs do not correspond to physical systems, which have boundaries.

\subsection{Bethe Boundary Conditions}
For a Bethe boundary, Poisson's equation allows us to write
the small $k$ behavior of the charge density as
\begin{equation}
\frhoq_0(\kb)\sim -\frac{k^2}{4\pi}\phiB.
\end{equation}
The charging free energy is obtained by performing the integration in Eq.~\ref{eq:charging} to yield
\begin{equation}
\Delta G^c(Q) = -\frac{Q^2}{l\sqrt{2\pi}}\para{1-\frac{1}{\epsilon}}+Q\phiB.
\end{equation}
Therefore, the contribution arising from the boundaries is non-zero and given by
\begin{equation}
\Delta G^{c}_{\rm PB}=Q\phiB
\end{equation}
for Bethe boundaries.
Physically, this illustrates that the mere presence of a boundary with a vacuum,
even without any structural response at that boundary,
contributes to the charging free energy.
The inclusion of the Bethe potential in the estimate of the free energy induces a fundamental asymmetry
with respect to the sign of the ionic charge distribution, which arises solely from the boundaries
and is independent of the induced structure component of the charge free energy, $\Delta G^c_{\rm IS}$,
as shown in Fig.~\ref{fig:fes}a.
For a Gaussian charge distribution in the SPC/E model of water, $\Delta G^c_{\rm IS}(Q)$ is symmetric with respect
to the sign of $Q$.
The inclusion of a Bethe or physical boundary results in a charge hydration asymmetry arising solely from
boundary contributions.
This asymmetry is dependent on the sign of $\phiB$, which is typically negative for classical point charge models
but positive (and large) for a quantum mechanical descriptions of water~\cite{Leung:2007,Lueng:2009,Remsing_JPCL_2014,Kathmann:2011}.
However, explicit knowledge of the boundary contributions to the charging free energy enables
the removal of $\phiB$ contributions to facilitate comparison of the structural response of classical and quantum
models to charging, which have been shown to be similar for reasonable models of water~\cite{Remsing_JPCB_2018}.

\subsection{Distant Structural Boundary Conditions}

\begin{figure*}[tb]
\begin{center}
\includegraphics[width=0.99\textwidth]{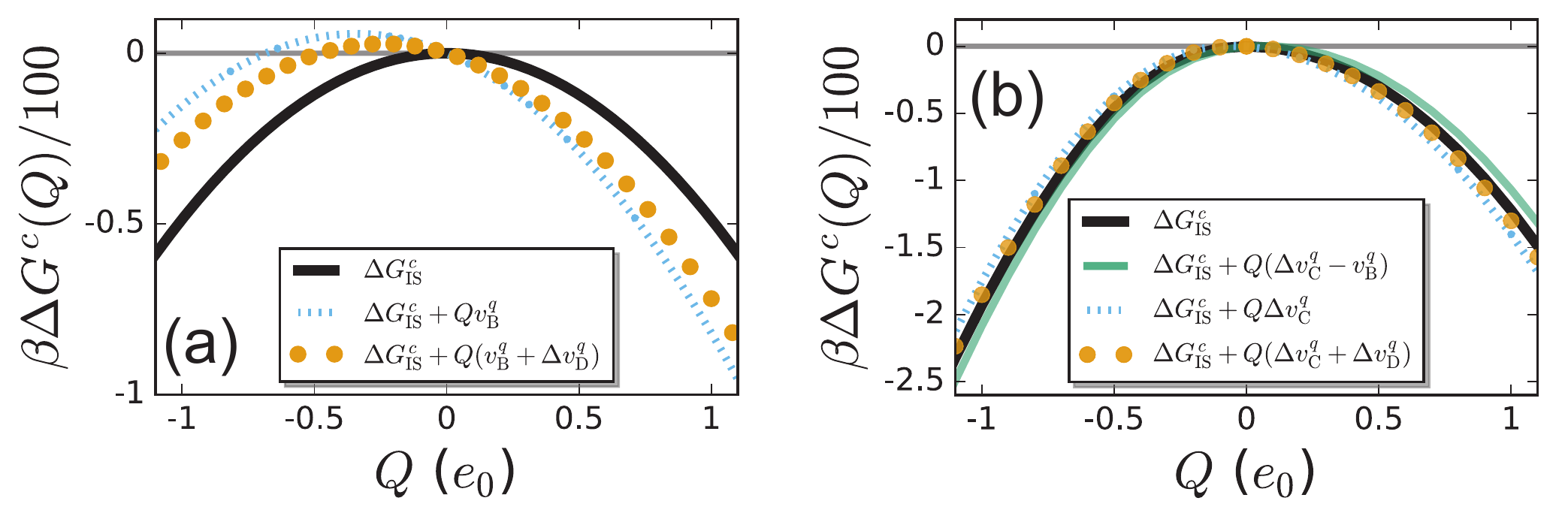}
\end{center}
\caption[Free Energies]
{
Charging free energies for (a) a Gaussian charge distribution of width $l=4.5$~\AA, as predicted by linear response theory,
and
(b) a point charge distribution inside a hard sphere solute excluding SPC/E water from a radius of roughly 2.6~\AA,
determined in previous work~\cite{Remsing_JPCB_2018}.
In each case, we show the induced structure component of the charging free energy, $\Delta G^c_{\rm IS}(Q)$,
as well as the impact of the preexisting boundary conditions discussed here.
Note that the hard sphere solute (b) exhibits both local and distant preexisting boundaries,
and the green curve illustrates the effect of local boundaries, $\Delta G^c_{\rm LPB}(Q)$,
on the charging free energy.
Here, we take $v^q_{\rm B}=-0.86$~V, $\Delta v^q_{\rm D}=+0.26$~V~\cite{Remsing_JPCL_2014},
and $\Delta v^q_{\rm C}=-0.46$~V.
}
\label{fig:fes}
\end{figure*}

The general formalism for describing structural boundaries is completely analogous to that for Bethe boundaries.
However, the potential difference across such a boundary is $\Delta v^q = \phiB + \phiD$.
In this case, the charge density of the bulk solvent is
\begin{equation}
\frhoq_0(\kb)\sim -\frac{k^2}{4\pi}\para{\phiB + \phiD}
\end{equation}
for small $k$.
The solvation free energy of a Gaussian charge in the presence of a distant structural boundary is then given by
\begin{equation}
\Delta G^{c}(Q) = -\frac{Q^2}{l\sqrt{2\pi}}\para{1-\frac{1}{\epsilon}}+Q\para{\phiB+\phiD}.
\end{equation}
The contribution from preexisting, distant structural boundaries is
\begin{equation}
\Delta G^{c}_{\rm PB}=Q\para{\phiB+\phiD},
\end{equation}
which includes a contribution from the interfacial structural response of the solvent, $\phiD$.
This contribution is decoupled from the charging of the solute and can be estimated by
studying the isolated physical interface.
Inclusion of $\phiD$ into the Gaussian charging free energy estimates in SPC/E water
partially cancels the effect of $\phiB$, as shown in Fig.~\ref{fig:fes}a.

\subsection{Implications for Ions with Excluded Volume Cores}
Typical classical ion models consist of an excluded volume and a point charge.
This state can be reached from the solvated Gaussian charge distribution by inserting an excluded volume
and subsequently shrinking the Gaussian charge, $l\rightarrow0$.
Carrying out this process affects both components of the charging free energy.
The linear Born theory estimate of $\Delta G^c_{\rm IS}(Q)$ can be obtained by
making the substitution $l \rightarrow R_{\rm B}\sqrt{2/\pi}$ in Eq.~\ref{eq:genFE}~\cite{Remsing_JPCB_2016}.
However, core insertion and shrinking of the Gaussian charge distribution are generally
non-linear processes that require explicit simulations to accurately determine $\Delta G^c_{\rm IS}(Q)$~\cite{Duignan2017,Remsing_JPCL_2014,Duignan2017a,Remsing_JPCB_2016,Remsing_JPCB_2018}.

This excluded volume introduces a local structural boundary around the point charge~\cite{Remsing_JPCL_2014,Remsing_JPCB_2016,Remsing_JPCB_2018},
which also modifies the preexisting boundary contribution to the charging free energy, $\Delta G^c_{\rm PB}(Q)$.
The boundary term in this case consists of two contributions, 
\begin{equation}
\Delta G^c_{\rm PB}(Q) = \Delta G^c_{\rm LPB}(Q) +\Delta G^c_{\rm DPB},
\end{equation}
where the first term arises from the \textit{local preexisting boundary}.
The second term is the contribution from \textit{distant preexisting boundaries}, discussed in previous sections,
and is unchanged by core insertion and shrinking the Gaussian charge.

Because the excluded volume core is a structural boundary, it can be readily evaluated following the discussion in the previous section,
\begin{equation}
\Delta G^c_{\rm LPB}(Q) = Q\para{\phiC - \phiB},
\end{equation}
where $\phiC$ is the physically relevant cavity interface component arising from solvent
structure~\cite{Remsing_JPCL_2014}.
Inclusion of $\Delta G^{c}_{\rm LPB}(Q)$ shifts the charging free energy in a manner that further makes
anion solvation free energies more favorable than that of cations, see Figure~\ref{fig:fes}b.

When system boundaries are taken into account, the Bethe potential contribution cancels between the two boundaries.
For example, when a distant Bethe boundary is present, 
\begin{equation}
\Delta G^c_{\rm PB}(Q)=Q\phiC,
\end{equation}
and the presence of a distant structural boundary adds a contribution of $Q\phiD$ to this result.
Both cases are shown in Figure~\ref{fig:fes}b, which shows that inclusion of distant boundaries, both Bethe
and physical boundaries, oppose the usual charge hydration asymmetry in the case of SPC/E-like water models
because both $\phiC$ and $(\phiC+\phiD)$ are negative.
This is in agreement with recent results from Cox and Geissler, who computed ionic charging free energies
in systems with and without an explicit water-vapor interface, and showed that proper inclusion of the boundary
potentials, as well as important system size and geometry-dependent dielectric response corrections, bring the
two sets of free energies into agreement~\cite{CoxGeissler}.

The cancellation of the Bethe potential across the two boundaries naturally arises within our structural
perspective.
In Fig.~\ref{fig:droplet}e, we show the electrostatic potential obtained for a system with an excluded volume fixed at the origin and a Bethe boundary far away.
As one moves from the vacuum toward the origin, the potential drops at the Bethe boundary by $\phiB$ and
remains constant throughout the bulk.
As the excluded volume is approached, oscillations appear in $\V(r)$ due to structural ordering around the solute,
before a sharp transition at the solute surface to a new constant value inside the core.
The resulting electrostatic potential inside the excluded volume core is $\phiC$.
Because there is a distant Bethe boundary present,
the cavity interface component, $\phiC$, is appropriately referenced to vacuum and independent of $\phiB$.
This leads to similar physically meaningful results for both classical and quantum models,
despite the large differences in their respective Bethe potentials~\cite{Remsing_JPCL_2014}.

For the solvation of neutral ion clusters, any contribution to the free energy from distant structural boundaries, \textit{e.g.} $Q\phiD$, cancels among the ions, because it is independent of their molecular structure.
In contrast, $\phiC$ depends on the size and shape of the ion core.
Therefore, its contribution to the total solvation free energy of neutral ion combinations only vanishes in the idealized situation where all ion cores are identical
and generally contributes to experimentally measured ion solvation thermodynamics.

\begin{figure*}[tb]
\begin{center}
\includegraphics[width=0.99\textwidth]{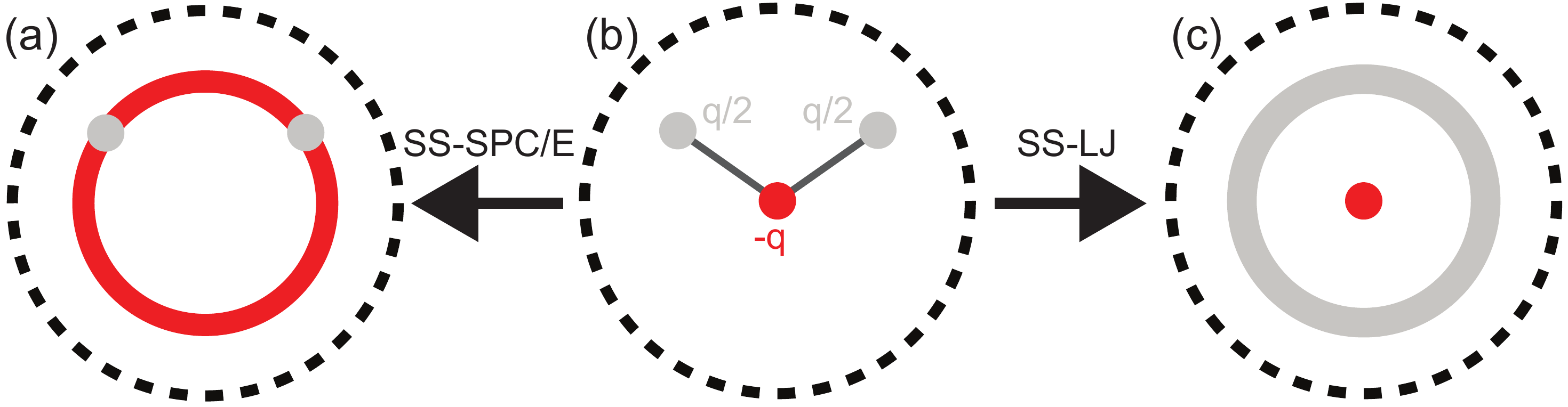}
\end{center}
\caption[Smeared Shell Models]
{
(a) The smeared shell variant of the SPC/E water model (SS-SPC/E) is obtained from (b) the original SPC/E model
by smearing the oxygen charge (red) onto a spherical shell of radius $r_{\rm OH}=1$~\AA \ (red in a).
Analogously, (c) the SS-LJ model is obtained by smearing the hydrogen site charges (gray) into a spherical shell
of radius $r_{\rm OH}$ (gray in c).
Black dashed circles indicate the LJ length scale of $\sigma=3.166$~\AA \ for the SPC/E water model.
}
\label{fig:smear}
\end{figure*}

To summarize, for model ions with their charge localized within excluded volume cores, $\phiB$
does not contribute to the electrostatic potential inside the solute core
when system boundaries are accounted for, such that the electrostatic potential has been appropriately referenced to the vacuum.
Consequently, the solvation free energy of a single ion of this type is independent of the bulk electrostatic (Bethe) potential.
In contrast, the Bethe potential does contribute to solvation thermodynamics in the absence of a solute core.

\section{Smeared Shell Models}

To further clarify the role of the Bethe potential (or lack thereof) in solvation thermodynamics,
we introduce and study extensions of the \emph{smeared shell} (SS) models introduced
in our previous work~\cite{Remsing_JPCL_2014}.
These models further illustrate the above points regarding electrostatic potentials and solvation thermodynamics
on a physical basis that we hope will clarify additional subtleties.
We use these models to show that one can systematically alter the Bethe potential without changing intermolecular forces,
and, consequently, solvation thermodynamics in most relevant cases.
Consider a three-site classical point charge SPC/E-like model of water~\cite{SPCE}, as shown in Fig.~\ref{fig:smear}b.
This model consists of a central oxygen site, at which a LJ potential and a point charge of magnitude $-q$
are located.
In addition, there are two hydrogen point charges of magnitude $q_{\rm H}=q/2$ located a fixed distance
$r_{\rm OH}$ from the center of the molecule, at a fixed H-O-H angle of $\theta$.

These fixed point charge models give rise to H-bonding through frustrated charge pairing~\cite{Remsing_JStatPhys_2011},
wherein point charges of opposite sign, located well within their respective repulsive molecular cores,
are strongly attracted to each other when neighboring molecules have the proper orientation.
However, this strong charge pairing attraction is opposed by the harshly repulsive forces arising from overlap of
the molecular LJ cores, which frustrates the charge pairing at typical inter-water distances,
resulting in an accurate yet simple model of H-bonding.
From this simple point charge model, we can construct two SS models. 
The first SS model we consider is obtained by smearing the oxygen point charge onto the surface of a sphere of radius $r_{\rm OH}$,
overlapping with the hydrogen point charges, resulting in the smeared shell SPC/E (SS-SPC/E) model
shown in Fig.~\ref{fig:smear}a.
Because the intramolecular charge distribution is well within the repulsive core, the intermolecular energies and
forces are unchanged by this smearing, as follows from Gauss's law, and consequently the liquid structure is unchanged.
The frustrated charge pairing picture of H-bonding persists after smearing in this fashion.
Therefore, the work required to charge a solute, with the charge distribution localized within an excluded volume,
is the same as that of the original SPC/E model,
and $\Delta G^c_{\rm IS}(Q)$ is unchanged upon smearing.
However, smearing the charge distribution in this manner results in $\phiB=0$,
removing the bulk component of the pre-existing boundary
contribution to the free energy~\cite{Remsing_JPCL_2014}, $\Delta G^c_{\rm PB}(Q)$.
This suggests that $\phiB$ cannot contribute to physical ion solvation thermodynamics.
In contrast, $\phiD$ is unchanged upon smearing and
contributes to single ion solvation thermodynamics when a distant structural boundary is present.
We also consider a SS model resulting from an analogous smearing of the hydrogen point charges,
resulting in a model with a spherical shell of positive charge of radius $r_{\rm OH}$ and a negative point charge
at its center, resulting in the smeared shell Lennard-Jones (SS-LJ) model shown in Fig.~\ref{fig:smear}c.
The resulting spherical charge distribution is well inside the repulsive LJ core.
This smearing has removed the strong interactions involving localized hydrogen point charges,
which compete with the harsh LJ repulsive core to create the frustrated charge pairing picture
of H-bonding in the full SPC/E water model.
Therefore, intermolecular interactions in a liquid composed of this model are due to the LJ potential alone,
and the resulting structure is identical to that of a simple LJ fluid.

In the presence of a structural boundary, $\phiD$ is non-trivially modified because of the
change in the solvent structural response.
In contrast, the Bethe potential,
$\phiB$, remains unchanged and is equal to that of the full SPC/E water model.
This model solvent does not respond to charging a solute whose charge distribution is inside an excluded volume,
such that $\Delta G^c_{\rm IS}(Q)=0$.
Therefore, there is no work performed (or required) throughout the solute charging process with this model.
%

\subsection{Generalized Bulk Model}
We now generalize the SS-LJ model of Fig.~\ref{fig:smear}c, and 
consider a system of such molecules, which each consist of a single, large LJ core.
Each LJ core has a point charge of magnitude $-q<0$ placed at its center, and a neutralizing
charge of opposite sign is smeared over a concentric spherical shell of radius $R_{\rm S}$, which is significantly smaller
than $\sLJ/2$.
As discussed above, the forces exerted between molecules are such that the internal charge distributions never overlap in any
relevant configuration of the system.
We additionally consider the presence of distant wall potentials,
which define the system volume $V$, outside of which there are no molecules.
By definition the charge densities never overlap, so we can readily write down the electrostatic potential
at any point in configuration space,
\begin{equation}
\V(\rb) = \frac{q}{R_{\rm S}}- \frac{q}{\len{\rb-\rb_i}}, {\rm \ for} \ \rb \ {\rm inside \ shell \ of \ molecule} \ i
\end{equation}
and $\V(\rb)=0$ otherwise, where $\rb_i$ is the position of molecule $i$.
The constant $q/R_{\rm S}$ ensures that the potential is continuous across the shell boundary.
We can then use $\V(\rb)$ to evaluate the ensemble averaged charge densities and electrostatic potentials exactly for this SS-LJ system.
%

\subsection{Simplified Description of Slab-like Perturbations}

\begin{figure*}[tb]
\begin{center}
\includegraphics[width=0.99\textwidth]{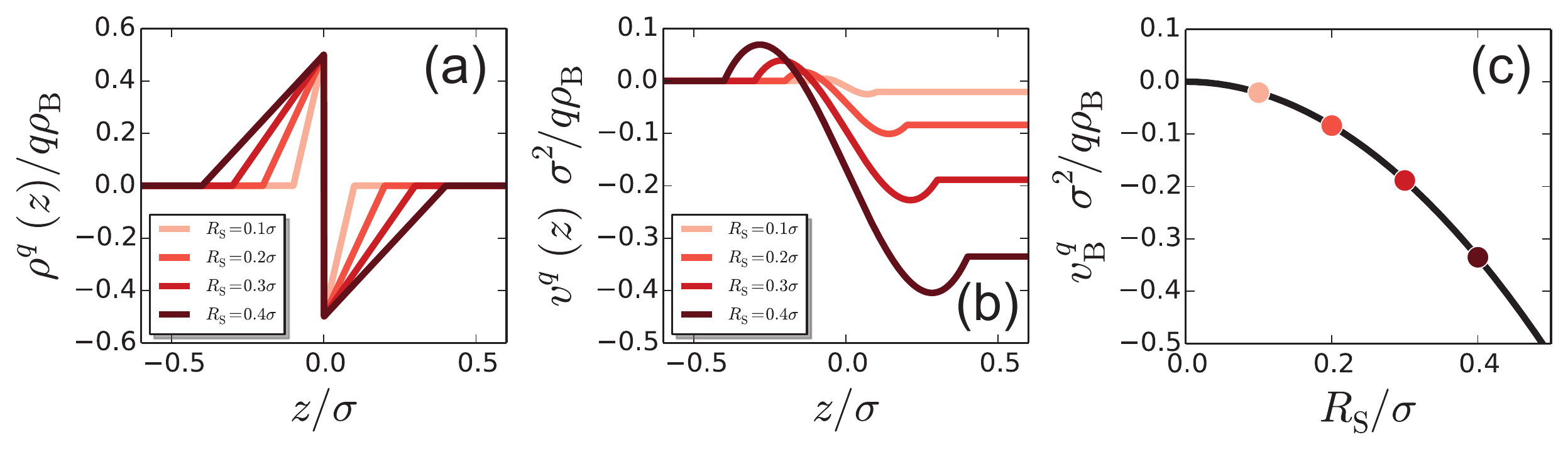}
\end{center}
\caption[Electrostatic Properties of SSLJ Model]
{
(a) Charge density, $\rho^q(z)$, (b) electrostatic potential, $\V(z)$, and (c) Bethe potential, $\phiB$,
for the Lennard-Jones smeared shell model, assuming a step-function form for the nonuniform density
in the presence of a hard wall located at $z=0$, which excludes density from the region $z<0$.
Data points in (c) correspond to $\phiB$ for the values of $R_{\rm S}$ shown in (a) and (b).
}
\label{fig:stepdens}
\end{figure*}

%
We now consider perturbing this model system by a hard wall located at $z=0$ and
oriented perpendicular to the $z$-axis.
For simplicity, we assume that the nonuniform number density near the hard wall is 
$\rho_{\rm LJ}(z)=0$ for $z<0$ and $\rho_{\rm LJ}(z)=\rhob$ for $z>0$,
\ie \ the density is a step function centered at $z=0$.
Under most conditions, the hard wall will induce a non-trivial, layered structure at the interface,
and $\rho_{\rm LJ}(z)$ introduced here is a simple approximation to the more complex transition from
zero to bulk density as one moves from inside the wall into the bulk liquid. 
However, one could also consider applying an additional one-body potential to the system
that yields the step-function density profile of interest in a more physical manner.
Such a boundary is equivalent to a Bethe boundary in this system.
The nonuniform charge density can be readily evaluated according to
\begin{equation}
\rhoq(z)=\int d\sb \rho_{\rm LJ}(z_s+z)\brac{q \frac{\delta(s-R_{\rm S})}{4\pi R_{\rm S}^2}-q\delta(\sb)},
\end{equation}
where $z_s$ is the $z$-component of $\sb$.
Performing the integration for the above form of $\rho_{\rm LJ}(z)$ yields
\begin{equation}
\rhoq(z) = \left\{ \begin{array}{ll} \sgn(z)\frac{ q \rhob }{2}\brac{\frac{\len{z}}{R_{\rm s}}-1}, \ \ \  -R_{\rm s}\le z \le R_{\rm s}\\
0, \ \ \ \ \ \ \ \ \ \ \ \ \ \ \ \ \ \ \ \ \ \ \ \ \ \ \ \ \ \ \ \ {\rm otherwise} 
\end{array} \right.
\label{eq:stepdens}
\end{equation}
This charge density is shown in Figure~\ref{fig:stepdens}a for several values of $R_{\rm S}$,
which illustrate that the deviation in the average charge density from zero is localized to the interface.
Equation~\ref{eq:stepdens} serves to illustrate that even in this simple model system there is a nonzero
charge density induced at an idealized hard wall boundary. 
The electrostatic potential can also be evaluated by integrating the charge density. 
This yields
\begin{equation}
\V(z) = \left\{ \begin{array}{lll} 0, \ \ \ \ \ \ \ \ \ \ \ \ \ \ \ \ \ \ \ \ \  z<-R_{\rm S} \\
f(z) \phiB, \ \ \ -R_{\rm S}<z<R_{\rm S} \\
\phiB, \ \ \ \ \ \ \ \ \ \ \ \ \ \ \ \ \ \ \  z>R_{\rm S}
\end{array} \right.
\end{equation}
where
\begin{equation}
f(z)=\brac{-\para{\frac{z}{R_{\rm S}}}^3 + \frac{3}{2}\para{\frac{z}{R_{\rm S}}} + \frac{1}{2} },
\end{equation}
which is consistent with the Bethe boundary nature of the system and shown in Fig.~\ref{fig:stepdens}b.
As with the charge density, variations in $\V(z)$ are localized to the interfacial region.
In this case, the electrostatic potential difference between the vacuum and the bulk is equal to the Bethe potential,
\begin{equation}
\phiB=-\frac{2}{3}\pi q \rhob R_{\rm S}^2.
\end{equation}
This potential can be readily tuned through the choice of $R_{\rm S}$, as shown in Fig.~\ref{fig:stepdens}c.
This tuning of the Bethe potential through $R_{\rm S}$ has no impact on the LJ intermolecular structure of the system,
as long as $R_{\rm S}<\sLJ/2$.

\subsection{Ion Solvation in the SS-LJ Model}

We now consider inserting a model ion into this bulk solution, far from the confining wall.
The model ion is composed of a LJ core with a point charge of magnitude $Q$ located at its center, such that the charge distribution
of the ion and the solvent do not overlap, and for simplicity, we take the limit that the ion-solvent LJ potential is equivalent
to the solvent-solvent LJ potential.
In this case, the solvent does not polarize in response to the ion, and the ion-solvent $g(r)$ must equal
that of the solvent-solvent $g(r)$.
Insertion of the ion is then equivalent to inserting another solvent molecule.
The solvation free energy of the ion is given by the chemical potential, $\mu$, of the LJ solvent,
$\Delta G=\mu$~\cite{WidomBook,WCA,TheorySimpLiqs}.
Note that $\mu$ does not involve the Bethe potential, $\phiB$.
The independence of $\Delta G$ with respect to $\phiB$ can be understood by noting that the system now has two sets of boundaries:
(i) the hard wall, across which there is a potential change $\phiB$, and
(ii) the ion core, across which the solvent electrostatic potential changes by $-\phiB$.
Similarly, ``real'' hydration free energies of single ions,
computationally estimated through appropriate reference of the electrostatic potential to the vacuum,
do not depend on $\phiB$, the constant electrostatic potential of the bulk phase.

Without appropriate reference to the vacuum, achieved here through the presence of the hard wall,
one would arrive at an ionic solvation free energy of $\Delta G=\mu - Q\phiB$.
In this case, the free energy unphysically depends on intramolecular charge densities
that have no impact on intermolecular forces and therefore the work required to solvate the ion. 
This situation is demonstrated by the calculations of Harder and Roux on ion solvation
in a rigid model of liquid methane~\cite{Harder:2008}.
Such a model is spherically symmetric on average
and excludes charge overlap between the intramolecular distribution of the solvent
and that of the ion, analogous to the SS-LJ model studied here. 
Thus, the free energy required to charge an ionic core in this model of methane should depend only on the magnitude
of the ion charge, e.g. it is symmetric with respect to $Q$.
However, Harder and Roux demonstrate that when using PBCs, without an appropriate boundary reference,
the difference is charging free energy between a cation and anion of the same is about 10~kcal/mol, 
while appropriate referencing to the vacuum yields the expected difference of zero;
for such a methane model, physical and Bethe boundaries are identical because there is not a dipole contribution.

\section{Conclusions}

In this work, we have illustrated the effects of system boundaries on the solvation free energies
of charged particles in dielectric media.
Appropriately referencing the electrostatic potential inside the condensed phase to
that in vacuum is of the utmost importance for the correct interpretation and calculation
of solvation free energies.
For the solvation of a single charged particle without an excluded volume core,
and in a system of finite size,
the potential of the bulk phase contributes to the solvation free energy,
through a term that is linear in the charge of the particle.
Such charge distributions probe the intramolecular charge distributions of solvent molecules,
and this charge overlap leads to this bulk, Bethe potential contribution to thermodynamics.
Charge distributions that are not confined within an excluded volume 
are used to experimentally probe
bulk potentials through electron holography, for example, by passing high energy electrons through samples of interest~\cite{McCartney_2007,Simon_2008,Ross_2015}.
These measurements yield average bulk phase potentials of 3-3.5~eV~\cite{Prozorov20170464}, consistent
with predictions from \emph{ab initio} simulations~\cite{Kathmann:2011}.
When a solute excluded volume core is present in a finite size system,
the presence of two boundaries, where one boundary defines the system size and the other the core region, ensures that
the potential of the bulk phase does not contribute to solvation thermodynamics.
The contribution from the Bethe potential, $\phiB$, cancels upon traversing
both boundaries.

The solvation free energy of physical ions involves
structural rearrangements of the bulk solvent that occur both at core boundaries near the 
solute and at distant  system boundaries.
Consequently, electrostatic potentials inferred from electrochemical measurements, roughly 0.1~eV or less,
are significantly smaller than those reported from electron holography.
Moreover, because the solute charge distribution cannot overlap with that of the solvent,
single ion solvation free energies are independent of the intramolecular charge distribution
of solvent molecules, such that classical and quantum simulations yield similar predictions.

Finally, we note that our work may aid in interpreting assumptions
made during the experimental determination of single ion solvation free energies.
In practice, one cannot directly measure a single ion solvation free energy, but only that
of a neutral collection of ions.
To decompose this free energy into single ion components, so-called ``extra-thermodynamic'' assumptions
must be made~\cite{Hunenberger2011}.
These hopefully physically reasonable assumptions enable the appropriate referencing of solvation free energies
to a standard,
and consequently the determination of solvation free energies of singles ions.
For instance, an often used extrathermodynamic assumption is the
tetra-phenyl arsonium and tetra-phenyl borate (TATB) hypothesis~\cite{Beck_2018,Duignan_2018,Hunenberger2011}. 
Within this framework, one can measure the combined solvation free energy of a large cation, TA, and a large anion, TB,
which have approximately the same size. 
It is also assumed that each ion contributes equally to the total solvation free energy,
which can then be split equally between TA and TB.

Our results indicate that the TATB assumption is invalid,
in agreement with the conclusions of recent simulation~\cite{Beck_2018,Duignan_2018}
and experimental~\cite{Scheu:2014aa} work,
and suggest alternative routes for determining appropriate references
for experimental measurements.
For two spherical ions of equal size --- an approximation of TA and TB ---
the total solvation free energy does not involve boundary terms and is given by
\begin{align}
\Delta G^{+-} &\equiv \Delta G(Q) + \Delta G(-Q) \\
&=\Delta G^c_{\rm IS}(Q)+\Delta G^c_{\rm IS}(-Q) \\
&< 2\Delta G^c_{\rm IS}(Q).
\end{align}
The inequality arises from the charge hydration asymmetry of ionic solvation,
$\Delta G^c_{\rm IS}(Q) > \Delta G^c_{\rm IS}(-Q)$; note that both quantities are negative~\cite{Beck_2018,Duignan_2018,Hunenberger2011,Remsing_JPCB_2016,Rajamani:2004,Mukhopadhyay:2012,Scheu:2014aa}.
Thus, the solvation free energies cannot be equally divided between the two ions, due
to this thermodynamic asymmetry, ultimately stemming from the asymmetry of the water molecule.
However, $\Delta G^c_{\rm IS}(Q)$ can be computed with little ambiguity from molecular simulations,
suggesting that accurate first principles simulations~\cite{Duignan2017a,SCANWater,SCAN-Ions}
may provide accurate estimates for the TA and TB solvation free energies,
which can then be used to reference experimental measurements.

We also note that single ion free energies will necessarily involve boundary terms,
which cancel when summing to obtain a neutral solution.
The difference in the single ion solvation free energies of a cation and anion of equal size
is given by
\begin{align}
\Delta \Delta G(Q) &\equiv \Delta G(Q)-\Delta G(-Q) \\
&=\Delta G^c_{\rm IS}(Q) - \Delta G^c_{\rm IS}(-Q) + 2Q\brac{\phiC+\phiD},
\end{align}
and is independent of the Bethe potential,
as expected from our discussions above.
The terms due to physical boundaries will formally contribute to the difference in single ion solvation
free energies between cations and anions of equal size,
but these boundary terms cannot be determined from the measurement
of solvation free energies of neutral collections of ions and alternative methods must be used to determine
their value.

We hope that these observations will aid in developing more accurate
extra-thermodynamic assumptions, possibly incorporating first principles simulation data,
as well as the development of thermodynamically-informed ion force fields for use
in molecular simulations~\cite{Netz:2012fk}.

\acknowledgements
This research was supported in part by NSF CHE-1300993.
We thank Chris Mundy, Greg Schenter, and Marcel Baer (Pacific Northwest National Laboratory),
Tim Duignan (University of Queensland),
Ang Gao (Massachusetts Institute of Technology),
and Teddy Baker (University of Maryland) for stimulating discussions.

\bibliographystyle{spphys}       

\end{document}